\documentclass[a4paper,11pt]{article}
\pdfoutput=1

\usepackage{aas_macros}
\usepackage{jcappub}

\usepackage[T1]{fontenc}
\usepackage{graphicx}
\usepackage{ctable}
\usepackage{comment}

\title{Predicting dust emission using galactic 21\,cm data}

\author[a,b]{Guangyu Zhang,}
\author[b]{Chi-Ting Chiang,}
\author[b]{Chris Sheehy,}
\author[b]{An\v{z}e Slosar}
\author[a]{and Jian Wang}

\affiliation[a]{State Key Laboratory of Particle Detection and
Electronics, University of Science and Technology of China, Hefei
230026, China}
\affiliation[b]{Physics Department, Brookhaven National Laboratory, Upton, NY 11973, USA}

\emailAdd{zguangyu@mail.ustc.edu.cn}
\emailAdd{ctchiang@bnl.gov}
\emailAdd{csheehy@bnl.gov}
\emailAdd{anze@bnl.gov}
\emailAdd{wangjian@ustc.edu.cn}

\abstract{Understanding large-angular-scale galactic foregrounds is
  crucial for future CMB experiments aiming to detect $B$-mode
  polarization from primordial gravitational waves. Traditionally, the
  dust component has been separated using its different frequency
  dependence. However, using non-CMB observations has potential to
  increase fidelity and decrease the reconstruction noise.  In this
  exploratory paper we investigate the capability of galactic 21\,cm
  observations to predict the dust foreground in intensity. We train a
  neural network to predict the dust foreground as measured by the
  Planck Satellite from the full velocity data-cube of galactic 21\,cm
  emission as measured by the HI4PI survey. We demonstrate that
  information in the velocity structure clearly improves the
  predictive power over both a simple integrated emission model and a
  simple linear model. The improvement is significant at arc-minute
  scales but more modest at degree scales. This proof of
  principle on temperature data indicates that it might also be
  possible to improve foreground polarization templates from the same
  input data.}

\begin{document}
\maketitle
\flushbottom

\section{Introduction}
\label{sec:intro}

Modeling and understanding the foregrounds contaminating Cosmic Microwave
Background (CMB) maps is crucial for deriving robust cosmological constraints
from them. So far, the two main approaches employed in reducing the impact
of unwanted foregrounds on the CMB maps rely on i) minimizing the problem
by choosing the cleanest parts of the sky and ii) projecting the unwanted
foreground components by the different spectral indices of various components
in the microwave sky to build maps of foregrounds.

As we are approaching the forthcoming generation of the CMB experiments,
exploring other possible techniques can be beneficial. This is particularly
important for the future measurements that aim to constrain the presence of
tensor modes in the primordial fluctuations, both for the ground-based
observations focusing on small areas of the sky (such as Simons Observatory \cite{Ade:2018sbj}
and CMB-S4 \cite{Abazajian:2016yjj}) and the satellite missions covering the
full sky (such as PICO \cite{Hanany:2019lle}, and LiteBIRD \cite{Hazumi:2019lys}).
Many of the future experiments will be incredibly deep, with sensitivities
below $1\,\mu$K\,arcmin, an over two order of magnitude increase over the
depth of Planck Satellite maps \cite{Akrami:2018vks}.
Therefore, a foreground that might have been completely negligible at
the Planck sensitivity level might suddenly be crucial. If these future
observations will be performed from the ground, the situation is even
more difficult as the range of frequency bands suitable for observation
is considerably smaller than in the space. Finally, the polarization
measurements have inherently more degrees of freedom as both the polarization
intensity as well as the orientation can, in principle, change with frequency,
an effect known as decorrelation \cite{Sheehy:2017gfx,Akrami:2018wkt}.
Therefore it would be helpful to develop additional methods for understanding
foregrounds.

The most important foregrounds at frequencies of interest from the
ground are the synchrotron and dust. The former has a strongly falling
spectrum and becomes very small at frequencies beyond $\sim
70$\,GHz. The latter has a rising spectrum and starts to dominate at
frequencies higher than $\sim150$\,GHz. While the CMB fluctuations
peak at around 220\,GHz, the global foreground minimum is at lower
frequencies of 70 -- 100\,GHz \cite{Ade:2015tva}.  In the
CMB data analysis, the raw temperature maps of the sky are first
cleaned using one of the component separation techniques (see e.g.
Ref.~\cite{Akrami:2018mcd} and reference there-in). The resulting CMB
maps often contain potentially significant levels of residual
foreground contamination. If a template for any such contamination is
known, it can be exactly marginalized in the estimation of the power
spectrum, a technique called \emph{template marginalization}. In
optimal quadratic estimation approaches, this is achieved by assigning
infinite variance to the linear combinations of pixels corresponding
to the template (see e.g Ref.~\cite{Elsner:2016bvs}).  In
pseudo-$C_\ell$ approaches the same effect can be achieved by fitting
and subtracting the template and then correcting for the small bias
this introduces in the power spectrum measurement (see
e.g. Ref.~\cite{Alonso:2018jzx}).  In both cases, the more faithful
the template map is, the better the results, but even with imperfect
templates, the biases in estimated power spectra can be reduced to
perhaps satisfactory levels. In particular, it is also possible to
marginalize over several templates, which in effect models the true
foreground contamination as some linear combination of those
templates. As the number of templates increases, the statistical power
is being lost, but template marginalization cannot introduce a bias,
\emph{unless} template correlates with the CMB signal. Therefore, it
is useful to consider methods that predict foregrounds based on data
that cannot be correlated with the true CMB signal. This lead us to
consider the galactic 21\,cm data.

The 21\,cm line is the hyperfine splitting of the ground level of neutral
hydrogen which arises due to alignment (or not) of the spins of electron
and proton. Galactic 21\,cm emission traces the neutral hydrogen in our
own galaxy. Since various components of the milky way (stars, dust, neutral
and ionized hydrogen, etc.) trace each other, it is reasonable to expect that
21\,cm measurements might be useful for predicting the dust foreground
contaminant. This is particularly attractive since the 21\,cm maps are
measured at frequencies which are orders of magnitude away from frequencies
relevant for observations of CMB fluctuations. We can therefore be absolutely
sure that these maps are blind to any structure in the CMB fluctuations.
Note that this is not true when one attempts to generate foreground maps
relying purely on the different spectral indices.

We are not the first ones to consider this idea. Several papers have
discussed how dust is traced by the integrated intensity in the 21\,cm
observations. Already in 1955, Ref.~\cite{1955ApJ...121..559L} has
noticed the association of dust and galactic 21\,cm line. More recently,
Ref.~\cite{Clark:2015cpa} has shown that the integrated intensity in
galactic 21\,cm data correlates with the dust polarization angle.
However, the 21\,cm dataset is considerably richer: the line is well
resolved by modern surveys giving the full velocity structure of the
hydrogen along the line-of-sight -- at each spatial position one can
measure the full profile of the velocity line. It is entirely plausible
that the full information about the velocity structure along the
line-of-sight encodes information about the physical environment in
which the neutral hydrogen exists and can therefore be useful in inferring
the dust polarization angle \cite{Clark:2018wyh,2019ApJ...874..171C,Clark:2019gap}.
In this paper we explore this possibility in a way that does not assume
any concrete physical model but uses existing data to see if dust can
be predicted from many maps corresponding to velocity slices in 21\,cm
measurement using modern machine-learning techniques. As a first exploratory
work we perform this using intensity alone, i.e. predicting intensity of
dust emission from the full velocity cube of 21\,cm data. The main result
is that there indeed seems to be more information the velocity slices
beyond that present in the integrated intensity map and which cannot
be modeled by a naive linear model of the type which is traditionally
utilized. This proof of concept anticipates future work involving
polarization data, but  it could also be useful itself for inferring
the true cosmic infrared background.

The question remains whether this method is useful at all. If one
needs the traditional component separation in order train the neural
network on one patch of the sky, then one might expect that this
correlation will never be useful as long as traditional component
separation is available. While we do not yet give a definite method
for doing that, the general point remains that if more information is
available, it can only be beneficial. For example, one could imagine a
fitting procedure that uses multi-frequency observations of the CMB
sky, together with galactic 21\,cm maps to simultaneously solve for
individual component maps \emph{and} the way these individual maps
map to the data vector: spectral indices and their variations, but
also the nonlinear and somewhat non-local relation between microwave
and 21\,cm dust emission. It might also be possible, for example, to
train on a small patch of extremely deep data microwave data and use
this training where only deep 21\,cm maps are available (in general,
these are signal dominated everywhere). The main point of this paper
is to show the existence of extra information.

The paper is structured as follows. In section \ref{sec:method} we
present the data used in this work, the models employed and general
caveats of the procedure. In section \ref{sec:results} we show the
result and the improvements that can be made in dust predictions over
what the integrated intensity predicts and finally conclude in section
\ref{sec:conclusions}. In appendix \ref{app:linear}, we demonstrate
the analytical calculation for the optimal weights of the linear model.
In appendix \ref{app:lambda} we study the effect of $\lambda$ values
on the correlation between the predicted and target maps.
In appendix \ref{app:split} we present the results for an addition sky
division for constructing the training and test datasets.

\section{Data and method}
\label{sec:method}

The main goal of this work is to investigate the capability of machine-learning
methods to predict the dust intensity from the galactic 21\,cm measurements.
For the target dust map we use the publicly available \texttt{COMMANDER}
dust component map\footnote{\url{https://pla.esac.esa.int}} with the resolution
of \texttt{Nside=512} produced by the Planck Satellite team using component
separation \cite{Akrami:2018mcd}. For the 21\,cm data we take the full-sky
measurements of the 21\,cm emission from our galaxy as measured by HI4PI
survey \cite{2016A&A...594A.116H}. HI4PI survey was constructed from the
Effelsberg-Bonn HI Survey \cite{2016A&A...585A..41W} and the Parkes Galactic
All-Sky Survey \cite{2015A&A...578A..78K}. The survey has angular resolution
of 16.2\,arcmin and sensitivity of 43\,mK, surpassing the legacy HI dataset
from the Leiden/Argentine/Bonn Survey \cite{2005A&A...440..775K}.

Because the standard neural network libraries are optimized to work with
data that are sampled on a Cartesian grid but not on the spherical coordinates,
we use \texttt{HEALPix}\footnote{\url{https://healpix.sourceforge.net}}
\cite{2005ApJ...622..759G} to gnomonically project 21\,cm and dust all-sky
maps to square cutouts. In particular, we divide the data into the north
galactic hemisphere used for training the model and south galactic hemisphere
used for measuring the effectiveness of the model. To ensure no overlap
between the training and test data sets, we do not use projection
centers with declination below 17.65 degrees.

\begin{figure}
\centering
\includegraphics[width=0.7\textwidth]{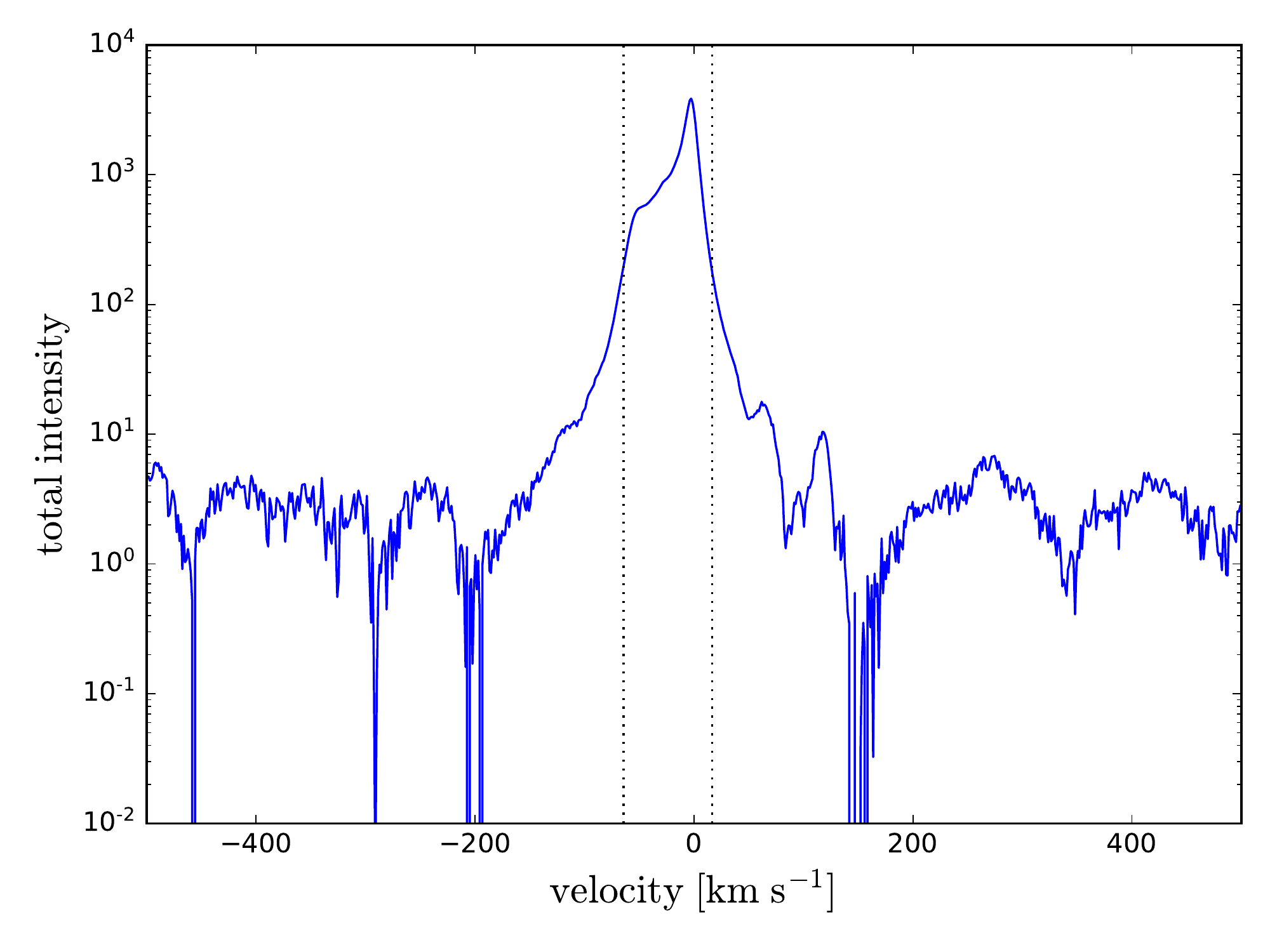}
\caption{Total intensity of 10000 random pointings (excluding areas covered
by the mask) for maps of all velocity slices. The vertical dotted lines
show the range of $-64.5\,{\rm km~s}^{-1}$ and $17\,{\rm km~s}^{-1}$, which
contains most of the 21\,cm signal.}
\label{fig:total_intensity}
\end{figure}

In detail, we process the data as follows:
\begin{itemize}
\item We start with full resolution maps that are pixelized on an
  \texttt{Nside=1024} \texttt{HEALPix} map.

\item For the 21\,cm galactic data, we use data in the velocity range
  between $-64.5\,{\rm km~s}^{-1}$ and $17\,{\rm km~s}^{-1}$ because it
  contains most of the 21\,cm signal (see figure~\ref{fig:total_intensity}).
  This leaves us with 64 individual maps $I_\nu$.\footnote{To further test
  the choice of the velocity range, we also train the neural network with
  data in the range of $\pm120\,{\rm km~s}^{-1}$, but we do not find a significant
  difference in the predicted correlation coefficients on all scales.}

\item We smooth data by applying a butterworth filter in $\ell$ domain
  with resolution $\ell < 460$. This is to prevent aliasing of small-scale
  power into large-scale power, but we do note that is a lossy process
  that could be sacrificing real information.
  
\item We apply the publicly available LR72 mask used in Ref.~\cite{Akrami:2018wkt}
  on the dust and 21\,cm maps for an effective sky fraction of 72\%.

\item We choose random cutout centers and rotations in the training and
  test datasets and project the input data onto a $64\times64$ Cartesian
  grid with the pixel size of 23.4 arcmin using a gnomonic projection routine
  in the \texttt{HEALPix} package. Such cutouts have a side length of 24.96
  degrees, corresponding to a fundamental mode expressed in units of standard
  spherical harmonic $\ell$ of $\ell_F=14.4$. To avoid regions largely covered
  by the mask, we discard the cutouts if the mask covers more than 70\% of
  the projections. Since we work with projected flat-space maps of a small
  area, the spherical harmonic decomposition can be trivially obtained using
  a two-dimension fast Fourier transform \cite{Bernardeau:2010ac}. We prepare
  the data so that the declination of the cutout centers follows $\cos(\rm DEC)$
  distribution to avoid oversampling the high-declination sky.
\end{itemize}

The result of this process is 50000 sets of maps from the north galactic
hemisphere for the training data and 1000 sets of maps from the south galactic
hemisphere for the test data. Each set contains a target dust map $T$ and
64 21\,cm maps $I_i$ applied with the same cutout and mask. Since the cutout
centers and rotations are randomly chosen, the 50000 maps are not completely
independent and some of them can overlap by chance. However, this should not
be an issue for training because the overlapping pixels will not appear at
identical locations. If the neural network just memorizes the relation between
dust and 21\,cm data for certain pixels, it will not be able to generalize
when these pixels show up in different locations for different cutouts,
and we do not observe such a problem.

We consider various predictive models that produce $\tilde{T}$ given input
maps $I_i$. Our success is characterized by the cross-correlation coefficient
given by 
\begin{equation}
\label{eq:crosscorr}
  r(\ell)=\frac{C_{T\tilde{T}}(\ell)}{\sqrt{C_{TT}(\ell) C_{\tilde{T}\tilde{T}}(\ell)}} \,.
\end{equation}
We compute the cross-correlation coefficients on individual Fourier modes
first and then bin the data into $\ell$ with bin size $\ell_F$. When compressed
into a single number, we average the cross-correlation coefficients of individual
Fourier modes over two bins: $\ell_F-100$ and $100-200$, denoted respectively
as $r_{50}$ and $r_{150}$. Note that $r$ is invariant under the total amplitude
rescaling of $\tilde{T}$ and hence we ignore any overall normalization factor.
A cross-correlation coefficient of 1 corresponds to perfect prediction $\tilde{T}=T$
on a given scale, while 0 a map that is completely uncorrelated. In principle
$r<0$ is possible, but these values correspond to a simple sign change.

We have attempted several models to predict $\tilde{T}$. In order of
increasing complexity, they are described below:

\subsection{Integrated intensity model}

Our baseline and simplest model is that the integrated 21\,cm intensity
traces the dust emission. That is, $\tilde{T}=\sum_i I_i$. Other models
will be judged by whether they offer (or not) improvement over this
simplest model.

\subsection{Linear combination model}
In the linear combination model, we model the target dust map as a
simple linear combination of input 21\,cm maps with a scale-independent
weight in Fourier space
\begin{equation}
 \tilde{T} = \sum_i w_i I_i \,,
\end{equation}
where the index $i$ runs from 1 to 64 for different velocity slices.
Note that the integrated intensity model is a special case of the
linear combination model with $w_i=1$. Appendix \ref{app:linear}
outlines the procedure for optimizing the linear weights given the
training data. In principle, the linear weights can be ``scale dependent'',
i.e. a distinct set of linear weights for each angular scale $\ell$.
However, our numerical experiment has shown that scale-dependent
weights are prone to overfitting. Namely, the results on \emph{test}
data get worse as the number of weights increases, indicating an overfitting
problem. Therefore, in this paper we restrict ourselves to scale-independent
weights.

\subsection{Deep neural network}

Our most advanced predictive model uses a deep neural network \cite{nature14539}.
Deep neural networks are a subset of machine learning algorithms used for
solving a wide range of problems, such as image recognition, machine
translation and image generation. A deep neural network which consists
of convolution layers is also called a convolutional neural network (CNN)
\cite{10.1007/978-3-319-10590-1_53,dumoulin2016guide}. A convolution layer
is some set of $N\times N$ kernel applied to the input. The kernel parameters
are adjusted by the back propagation algorithm during training. It can
effectively extract useful features from an image.

Our neural network structure is based on the U-Net model \cite{ronneberger2015u}.
U-Net is an image-to-image network, which takes images as input and also outputs
images. U-Net was designed to perform image segmentation tasks. The task in this
project is similar to image segmentation, i.e. both generating pixel-to-pixel maps.
Since this model has no fully connected layer, it can fit different input image
sizes using the same model.

\begin{figure}
\includegraphics[width=\textwidth]{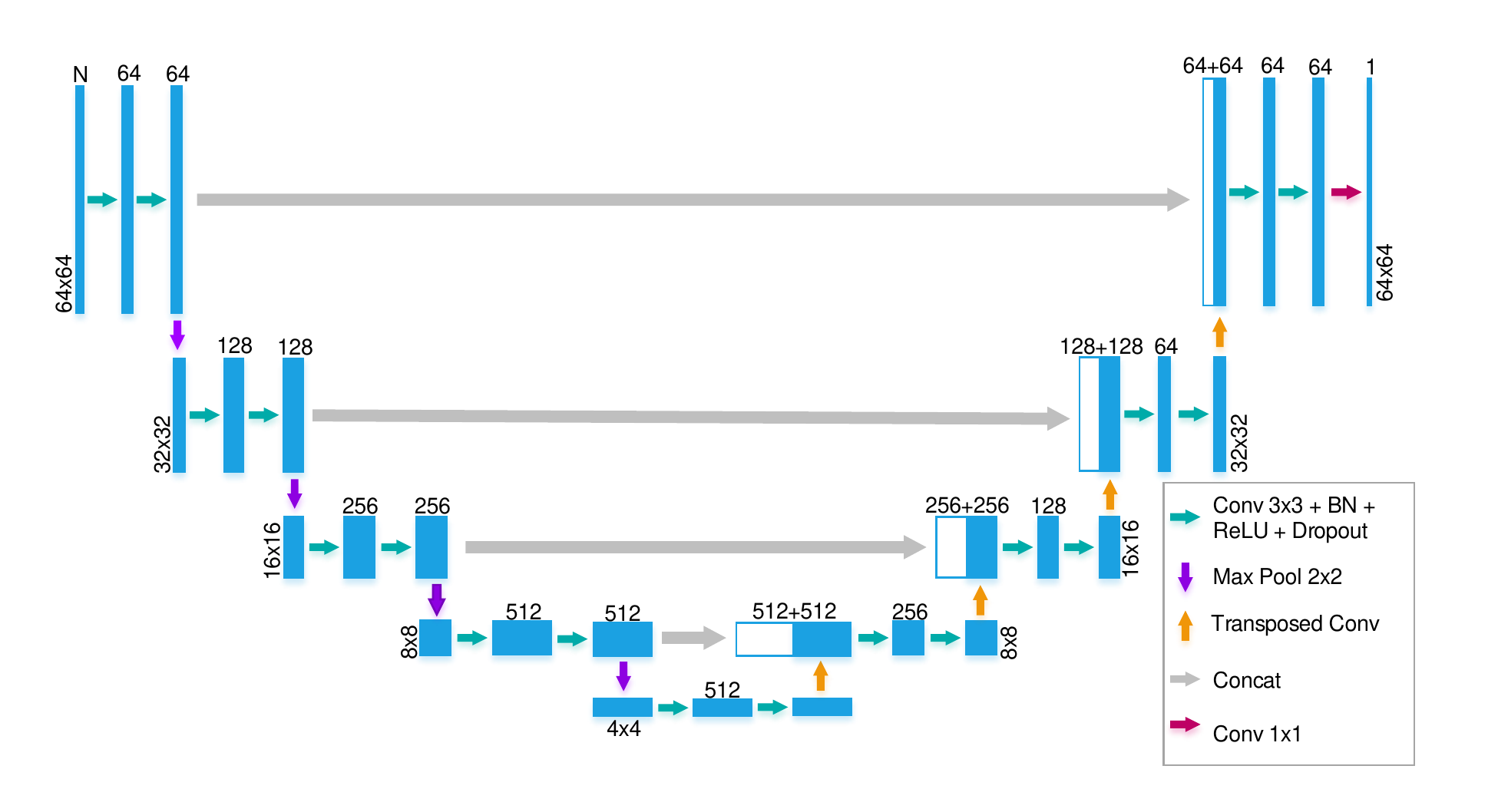}
\caption{The architecture of the deep neural network, U-Net, adopted in
this paper. The green right arrow contains a convolution layer with a
$3\times3$ kernel, stride 1 and same padding, a batch normalization layer,
a rectified linear unit for activation,  and a dropout layer with dropout
rate of 0.3; the purple down arrow represents a maximum pooling of $2\times2$
kernel and stride 2; the gray right arrow represents concatenation of
the current and the previous downward part outputs; the yellow up arrow
represents the transposed convolution with $2\times2$ kernel and stride 2.}
\label{fig:unet}
\end{figure}

Figure~\ref{fig:unet} shows the U-Net architecture with the $64\times64$
input images adopted in the paper. The U-Net has a downward part and an
upward part. The downward part consists of four sets of double convolution
operations and a maximum pooling (purple down arrow) with a $2\times2$
kernel and stride 2. Each convolution operation (green right arrow) contains
a convolution layer with a $3\times3$ kernel, stride 1 and same padding,
a batch normalization layer \cite{loffe43442}, a rectified linear unit
for activation, and a dropout layer with dropout rate of 0.3 to regularize
the network \cite{JMLR:v15:srivastava14a}. The upward part concatenates
the current and previous downward part outputs (gray right arrow) and
then uses transposed convolution \cite{odena2016deconvolution} with a
$2\times2$ kernel and stride 2 (yellow up arrow) to recover the original
image size. The concatenating operation keeps the pixel information along
with learned features. The output of the U-Net is images of $64\times64$
pixels, so they can be directly compared with the target dust maps.

We define the loss function of the neural network to be
\begin{equation}
 L(T,\tilde{T}) = \lambda\left[(1-r_{50})+(1-r_{150})\right] +
 \frac{1}{N_{\rm pixel}}\sum_{a\in{\rm pixel}}|T_a-\tilde{T}_a| \,,
\label{eq:loss}
\end{equation}
where $T$ and $\tilde{T}$ are respectively the true and predicted maps
and $\lambda$ is a hyper-parameter controlling the relative contributions
from the two terms of the loss function. The first term is the mean
cross-correlation coefficients (defined in eq.~\ref{eq:crosscorr}) at
angular scale bins $\ell=\ell_F-100$ and $\ell=100-200$. Since $|r(\ell)|\le1$,
minimization of $L(T,\tilde{T})$ leads to $r_{50}\to1$ and $r_{150}\to1$,
hence a higher correlation between $T$ and $\tilde{T}$. The second term
of the loss function is the mean pixel-by-pixel difference between $T$
and $\tilde{T}$. The pixel difference is quantified by the L1 loss instead
of the more popular L2 loss so that the model will not be too sensitive
to the outliers and bright spots, which occur when the point sources in
the dust and 21\,cm maps are not masked out. Though our primary goal is
to make the predicted maps be as highly correlated with the target maps
as possible, i.e. minimization of the first term, we include the pixel
difference in the loss function because the cross-correlation coefficient
is invariant under a total amplitude rescaling of $\tilde{T}$. This helps
break the degeneracy of the loss function for the neural network and the
training is more stable. In appendix \ref{app:lambda} we study the effect
of $\lambda$ values on the correlation between the predicted and target
maps. We find that for $\lambda\ge30$, the changes of the correlation is
not significant on all scales. In this paper we set $\lambda=30$ so that
the two loss terms have similar amplitudes (observed during training).

We train the model for 40 epochs, with a batch size of 64 so each epoch
has 782 iterations. The learning rate is set to $10^{-3}$ in the beginning
of the training, and is reduced by a factor of 10 for every 10 epochs.

\section{Results}
\label{sec:results}

\begin{figure}
\centering
\includegraphics[width=\textwidth]{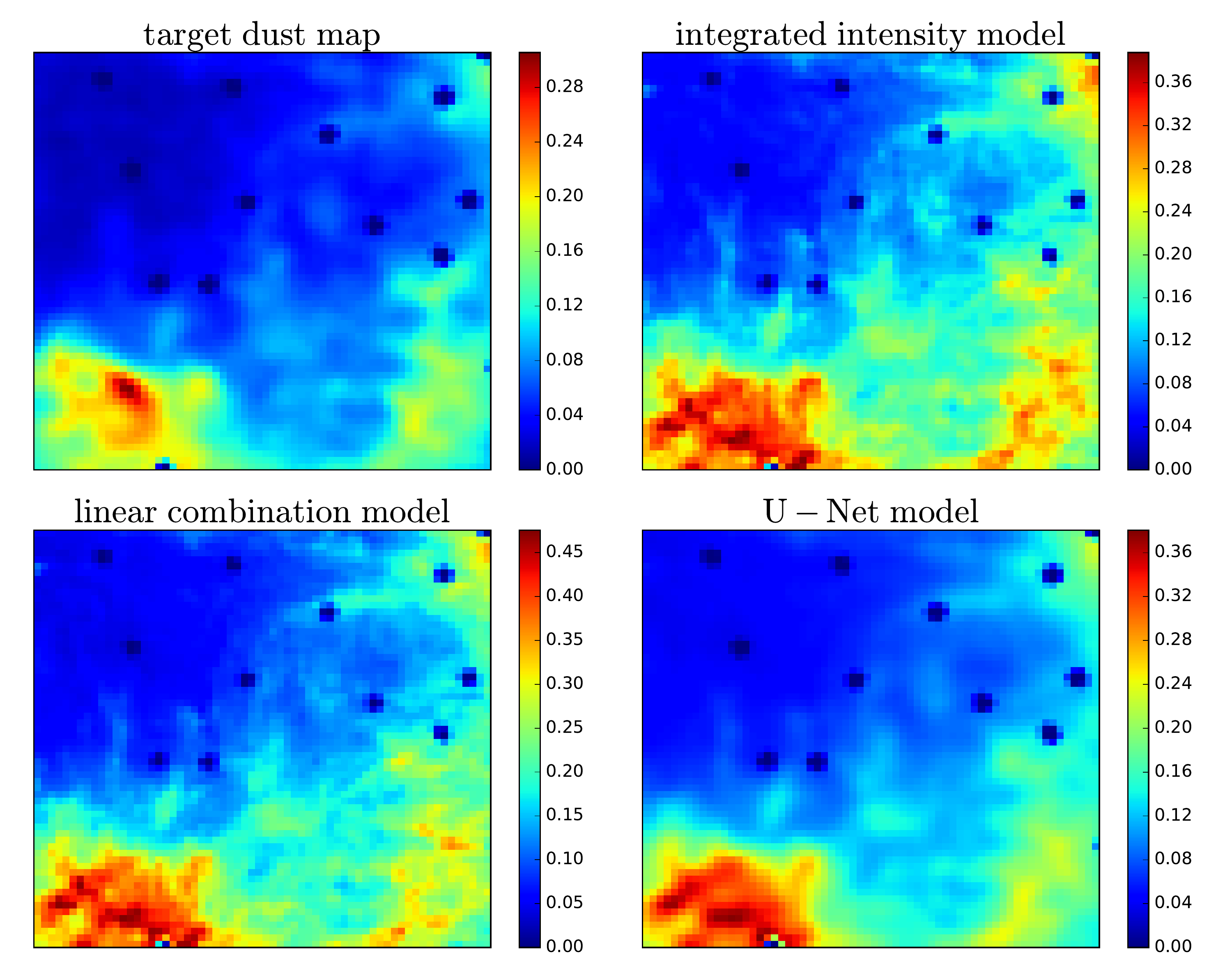}
\caption{Example maps of one cutout from the test set. (Top left) target
dust map; (top right) integrated intensity model map; (bottom left) linear
combination model map; (bottom right) U-Net model map. The darker dots are
the results of the mask. Since the normalization of the integrated intensity
map is different from the other maps, for a better comparison we normalize
the maps using their maximum pixel values. The ability of U-Net model to
improve the integrated intensity map on general morphological features in
the map is striking.}
\label{fig:maps}
\end{figure}

Figure~\ref{fig:maps} shows a set of example cutout maps from the test
dataset. The top left, top right, bottom left, and bottom right panels
show respectively the target dust map, the integrated intensity model
map, the linear combination model map, and the U-Net predicted map.
The darker dots are the results of the mask. Since the normalization
of the integrated intensity map is different from the other maps, for a
better comparison we normalize the maps using their maximum pixel values.
It is immediately striking how much better the neural network prediction
is compared to naive summing over all frequencies. However, we note that
visual comparison can often be very deceiving -- a human eye is trained
to pick individual features, which are dominated by high spatial frequency
features. For scientific interest we are most interested in recovery of
large-scale smooth components -- these are the scales most relevant for
the cosmological measurement such as tensor modes where we believe this
method could be most useful (once adapted for polarization).

\begin{figure}
\centering
\includegraphics[width=0.495\textwidth]{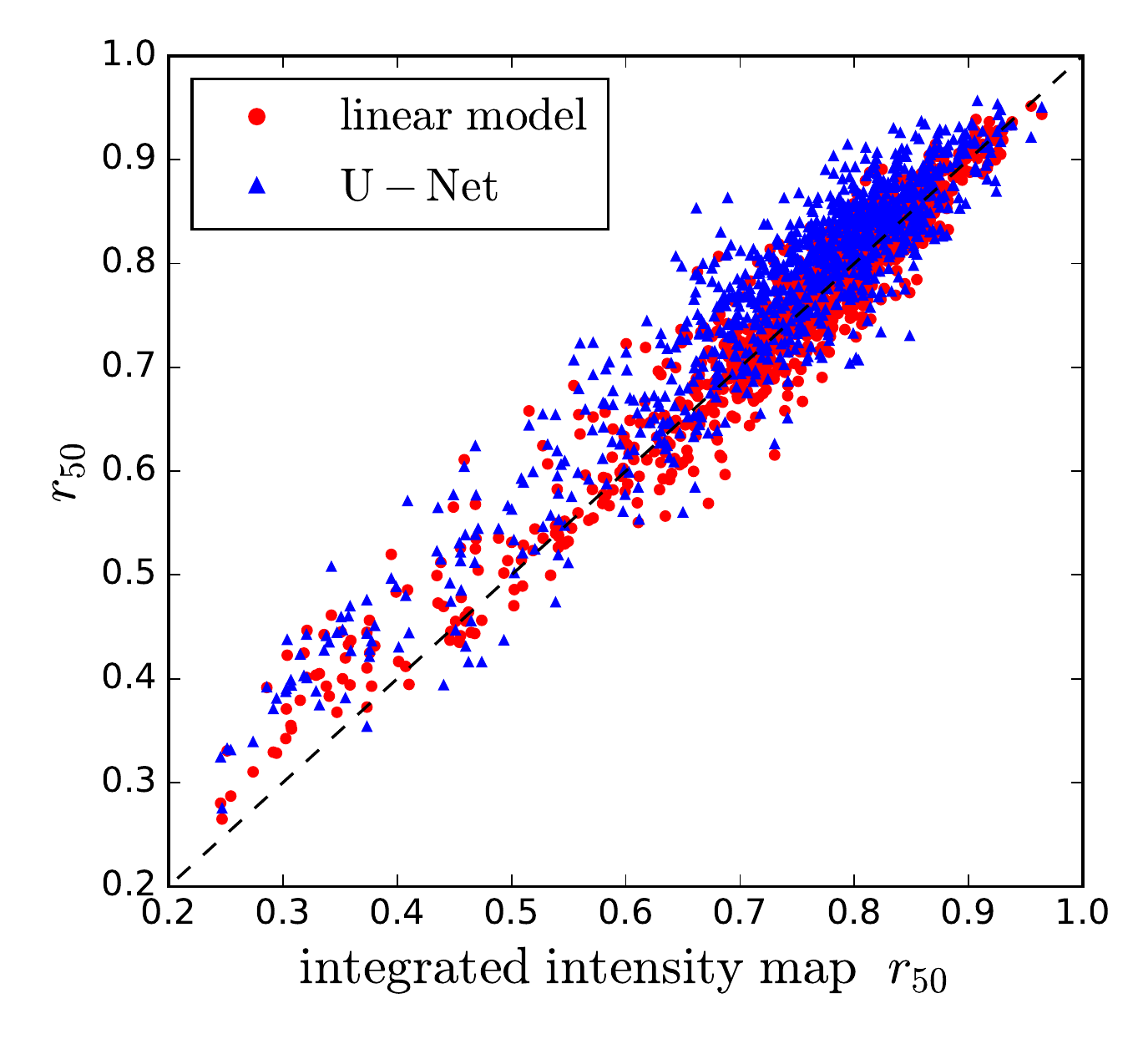}
\includegraphics[width=0.495\textwidth]{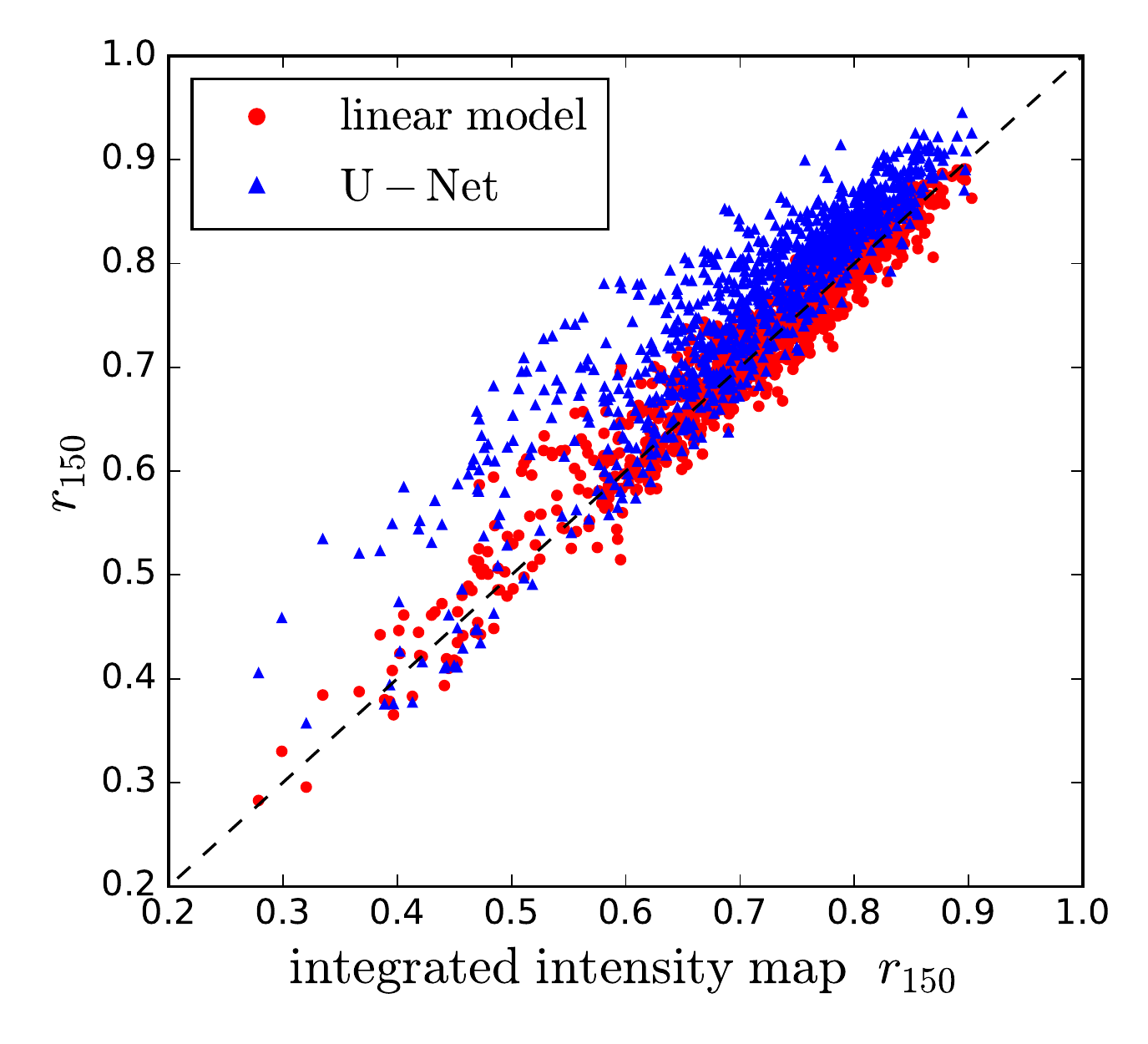}
\caption{Realization-by-realization comparison of $r_{50}$ (left) and
  $r_{150}$ (right) for the test set. The $x$-axis displays the
  cross-correlation coefficients between the target dust maps and the
  integrated intensity model maps, which we use as a reference to judge
  whether the other models provide improvement. The $y$-axis displays
  the cross-correlation coefficients between the target dust maps and
  the other models: the red circle is for the linear combination model
  and the blue triangle is for the U-Net model. Points above the dashed
  line indicate improvement, while points below correspond to deterioration
  with respect to the integrated intensity prediction. Out of 1000 realizations,
  U-Net outperforms the integrated intensity model in 764 and 889 realizations
  for $r_{50}$ and $r_{150}$, respectively.}
\label{fig:r50_r150}
\end{figure}

To better compare the performance between various models, we compute
the cross-correlation coefficients between the target dust maps and
maps predicted by different models. The left and right panels of
figure~\ref{fig:r50_r150} show a realization-by-realization comparison
from the test set of $r_{50}$ and $r_{150}$, respectively. The $x$-axis
displays the cross-correlation coefficients of the integrated intensity
model, which we use as a reference to judge whether the other models
provide improvement. The $y$-axis displays the cross-correlation coefficients
of the other models: the red circles are for the linear combination model
and the blue triangles are for the U-Net model. We first notice that the
linear combination model with the optimal weights computed from the
training set provides similar performance as the integrated intensity
model for both $r_{50}$ and $r_{150}$ on the test set. This indicates
that a linear transformation of the 21\,cm data is insufficient to outperform
the integrated intensity model. The U-Net model provides a nonlinear
transformation of the input 21\,cm data. We find that on both large
$(\ell_F\le\ell<100)$ and small $(100\le\ell<200)$ scales the U-Net model
outperforms the integrated intensity model for most of the realizations,
and the improvement is more apparent on small scale. Specifically, out
of 1000 realizations, U-Net outperforms the integrated intensity model
in 764 and 889 realizations for $r_{50}$ and $r_{150}$, respectively.
While the improvement is modest, it demonstrates that the neural network
indeed learns non-trivial features and transformation of the map compared
to the linear combination model.

\begin{figure}
\centering
\includegraphics[width=0.495\textwidth]{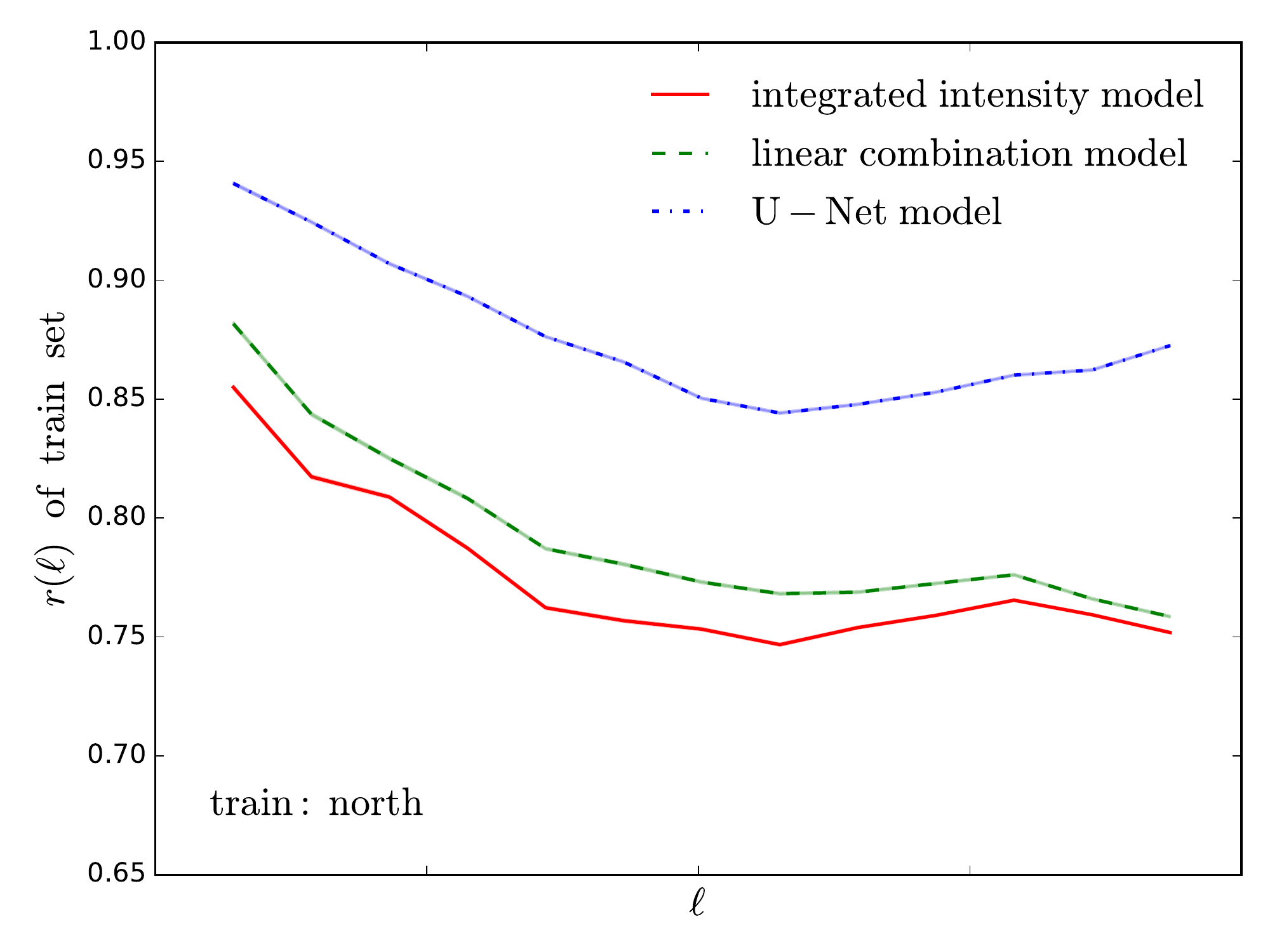}
\includegraphics[width=0.495\textwidth]{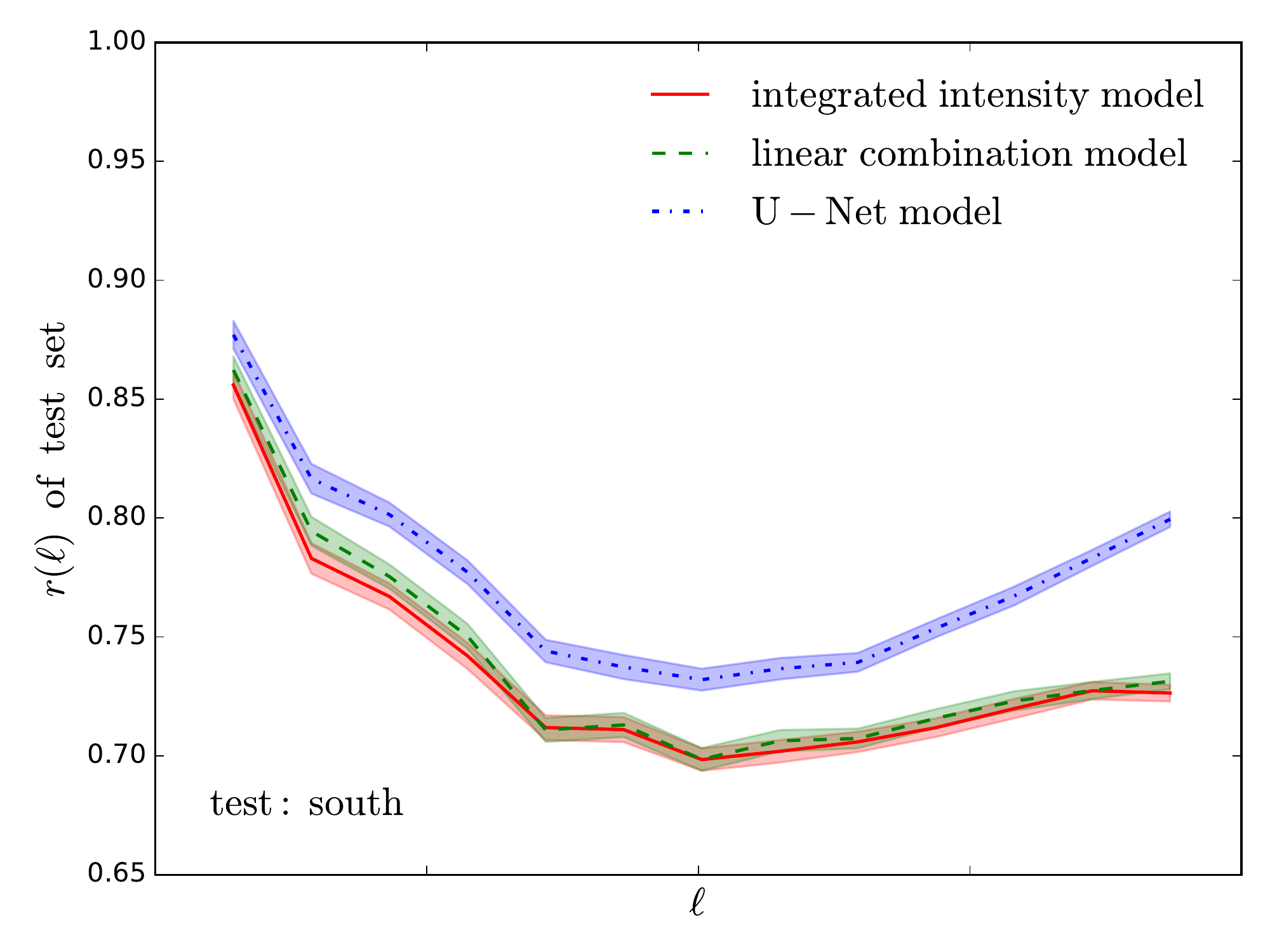}
\caption{Mean cross-correlation coefficient between the target dust maps
and different models as a function of $\ell$ for training (left) and test
(right) sets. The red solid, green dashed, and blue dot-dashed lines show
the results for integrated intensity, linear combination, and U-Net models,
respectively. The bands show the errors on the means.}
\label{fig:rl}
\end{figure}

To better examine the detailed performance of different models, in
figure~\ref{fig:rl} we plot the mean cross-correlation coefficient
between the target dust maps and different models as a function of
$\ell$ for training (left panel) and test (right panel) sets. The red
solid, green dashed, and blue dot-dashed lines show the results for
integrated intensity, linear combination, and U-Net models, respectively.
The bands show the errors on the means. For the test set, the conclusion
is the same as figure~\ref{fig:r50_r150}. Namely, the linear combination
model has a similar performance as the integrated intensity model on all
scales, whereas the U-Net model on average has a better performance on
all scales. Between training and test sets, we find that the linear and
U-Net models perform better for the training than the test set, and the
difference is larger for the U-Net model. This is a common outcome for
training a neural network and indicates modest overfitting. In particular,
the performance of the U-Net model worsens more strongly on large than
on small scales. This is probably due to the fact that there are fewer
large-scale modes and so it is easier to overfit on large scales. Another
interesting finding is that the integrated intensity model performs worse
for the test than the training set. We note that there are no parameters
to be fitted for the integrated intensity model, so there should be no
overfitting issue. In appendix~\ref{app:split}, we investigate this issue
further by dividing the sky into east and west galactic hemispheres, and
train and test on east and west hemispheres, respectively. However, we
do not find significant difference of the integrated intensity model.
Hence, we conclude that the difference of the integrated intensity model
performance between training (north) and test (south) sets shown in
figure~\ref{fig:rl} is likely a statistical fluctuation.

\section{Conclusions}
\label{sec:conclusions}

This paper describes the principle and the feasibility of predicting the
microwave dust signal using the 21\,cm hydrogen line signal. We compare three
models to do this work, an integrated intensity model, a linear combination
model, and a neural network model. The neural network model shows the best
performance on average but has a larger scatter than the linear combination
model.

More precisely, the neural network offers a modest improvement on all scales,
but the improvement is more apparent on small $(100\le\ell<200)$ than on large
scales $(\ell_F\le\ell<100)$. One possible explanation is that there are fewer
modes on large scales, hence it is more likely to overfit. Nevertheless, based
on the fact that the linear combination model on the full velocity slices of
21\,cm data cannot outperform the neural network, we conclude that the neural
network is indeed picking non-trivial features and transformation of the maps.

We have also attempted to train a neural network with moment maps
(i.e. maps of the type $\sum \nu^m I_\nu$) rather than full velocity
cubes, but this do not result in any improvements over the integrated
intensity map and hence we do not present those results. The same is
true using the integrated intensity map as the input to the neural network,
indicating that the resulting map is not just a sufficiently nonlinear
transformation of the integrated intensity map. However, it is possible
that a different network architecture coupled with preprocessing of
the velocity cube could result in improvements to the presented method.

The next step would be to apply this method to the
polarization. Polarization of dust is result of an alignment of dust
particles with the magnetic field. The distribution of the magnetic
fields is complex but again related to the morphology of matter
distribution in our galaxy. It has been shown that the integrated
intensity maps processed with a rolling Hough transform
\cite{Hough1962} correlate significantly with the dust polarization
angle \cite{Clark:2015cpa,Clark:2019gap}. However, the use of Hough transform in
that work was purely phenomenological. Moreover, the full velocity
cube can separate distinct streams of neutral hydrogen that appear
blended in projection and potential contain more information. Hence, a
natural extension of Refs.~\cite{Clark:2015cpa,Clark:2019gap} would be to replace the
integrated intensity maps with the full velocity cube and the Hough
transform with a neural network in order to \emph{learn} the necessary
transformation. We leave this for future work.

\acknowledgments
Results in this paper were obtained using the Institutional Cluster in
Scientific Data and Computing Center at the Brookhaven National Laboratory
under proposal 304400. Some of the results in this paper have been derived
using the \texttt{HEALPix} package.

\appendix
\section{Optimal weights for the linear model}
\label{app:linear}

Generally, the scale-dependent linear weights in Fourier space can be written as
\begin{equation}
  \tilde{T}_\ell = \sum_i w_{\ell,i} I_{\ell,i} \,.
\end{equation}
To solve the weights $w_{\ell,i}$, we minimize
\begin{equation}
 \chi^2_\ell = \sum_{n=1}^{N} \left( T^{n}_\ell - \tilde{T}^n_\ell \right)^2
 = \sum_{n=1}^{N} \left( T^{n}_\ell - \sum_i w_{\ell,i} I^n_{\ell,i} \right)^2 \,,
\end{equation}
where the superscript $n$ denotes the training set index and $N$ is
the number of training set. Requiring the derivative on $\chi^2_\ell$
with respect to $w_{\ell,j}$ to be zero, we obtain
\begin{equation}
 -2 \sum_{n=1}^{N} I^n_{\ell,j} \left( T^{n}_\ell - \sum_i w_{\ell,i} I^n_{\ell,i} \right) = 0 \,,
\end{equation}
which then leads to
\begin{equation}
 \sum_{n=1}^{N} \sum_i w_{\ell,i} I^n_{\ell,i} I^n_{\ell,j}
 = \sum_{n=1}^{N} T^{n}_\ell I^n_{\ell,j} \,.
\label{eq:chi2_sol}
\end{equation}
The above equation is a set of linear equations, and the solutions
can be found by performing one matrix multiplication. Specifically,
if we define the mean power spectra
\begin{equation}
 \bar{C}_{I_i I_j}(\ell) = \frac{1}{N} \sum_{n=1}^{N} I^n_{\ell,i} I^n_{\ell,j} \,, \quad
 \bar{C}_{T I_j}(\ell) = \frac{1}{N} \sum_{n=1}^{N} T^n_{\ell} I^n_{\ell,j} \,,
\end{equation}
we have
\begin{equation}
 \vec{w}_\ell = \left[\bar{\bf C}_{I_i I_j}(\ell)\right]^{-1} \vec{\bar{C}}_{T I}(\ell) \,,
\label{eq:wl_matrix}
\end{equation}
where $\vec{w}_\ell$ and $\vec{\bar{C}}_{T I}$ are vectors of 64
components and $\bar{\bf C}_{I_i I_j}$ is a $64\times64$ matrix.

Mathematically, one can show that the solution minimizing the
difference between $T$ and $\tilde{T}$ (equation (\ref{eq:wl_matrix}))
also maximize the correlation coefficient of the mean power spectrum
\begin{equation}
 \bar{r}(\ell) = \frac{\bar{C}_{T\tilde{T}}(\ell)}{\sqrt{\bar{C}_{TT}(\ell)\bar{C}_{\tilde{T}\tilde{T}}(\ell)}} \,,
\end{equation}
where
\begin{align}
 \bar{C}_{TT}(\ell) \:&= \frac{1}{N} \sum_{n=1}^{N} T^n_\ell T^n_\ell \,, \nonumber\\
 \bar{C}_{T\tilde{T}}(\ell) \:&= \frac{1}{N} \sum_{n=1}^{N} T^n_\ell \tilde{T}^n_\ell
 = \frac{1}{N} \sum_{n=1}^{N} T^n_\ell \sum_{i} w_{\ell,i} I^n_{\ell,i}
 = \sum_i w_{\ell,i} \bar{C}_{TI_i}(\ell) \,, \nonumber\\
 \bar{C}_{\tilde{T}\tilde{T}}(\ell) \:&= \frac{1}{N} \sum_{n=1}^{N} \tilde{T}^n_\ell \tilde{T}^n_\ell
 = \frac{1}{N} \sum_{n=1}^{N} \sum_{ij} w_{\ell,i} w_{\ell,j} I^n_{\ell,i} I^n_{\ell,j}
 = \sum_{ij} w_{\ell,i}w_{\ell,j} \bar{C}_{I_iI_j}(\ell) \,.
\end{align}
Specifically, taking the derivative on $\bar{r}(\ell)$ with respect
to $w_{\ell,k}$, we obtain
\begin{equation}
 \frac{\partial\bar{r}(\ell)}{\partial w_{\ell,k}}
 = \frac{\frac{\partial\bar{C}_{T\tilde{T}}(\ell)}{\partial w_{\ell,k}}\bar{C}_{\tilde{T}\tilde{T}}(\ell)
 -\frac{1}{2}\frac{\partial\bar{C}_{\tilde{T}\tilde{T}}(\ell)}{\partial w_{\ell,k}}\bar{C}_{T\tilde{T}}(\ell)}
 {\sqrt{\bar{C}_{TT}(\ell)\bar{C}^3_{\tilde{T}\tilde{T}}(\ell)}} \,,
\label{eq:rbar_deriv}
\end{equation}
where
\begin{align}
 \frac{\partial\bar{C}_{T\tilde{T}}(\ell)}{\partial w_{\ell,k}} \:&=
 \frac{1}{N}\sum_{n=1}^{N} T^n_\ell I^n_{\ell,k}
 = \bar{C}_{TI_k} \,, \nonumber\\
 \frac{\partial\bar{C}_{\tilde{T}\tilde{T}}(\ell)}{\partial w_{\ell,k}} \:&=
 \frac{1}{N}\sum_{n=1}^{N} 2 \sum_i w_{\ell,i} I^n_{\ell,i} I^n_{\ell,k}
 = 2\sum_i w_{\ell,i} \bar{C}_{I_iI_k} \,.
\end{align}
Combining the above results, the numerator in equation (\ref{eq:rbar_deriv})
becomes
\begin{equation}
 \sum_{ij} w_{\ell,i} w_{\ell,j} \left[ \bar{C}_{TI_k} \bar{C}_{I_iI_j}(\ell)
 - \bar{C}_{I_iI_k} \bar{C}_{TI_j}(\ell) \right] \,.
\end{equation}
Plugging in equation (\ref{eq:wl_matrix}), which is equivalent to
equation (\ref{eq:chi2_sol}) and
\begin{equation}
 \sum_i w_{\ell,i} \bar{C}_{I_iI_j} = \bar{C}_{TI_j} \,,
\end{equation}
into the above equation, we obtain
\begin{align}
 \:& \sum_{ij} w_{\ell,i} w_{\ell,j} \left[ \bar{C}_{TI_k}(\ell) \bar{C}_{I_iI_j}(\ell)
 - \bar{C}_{I_iI_k}(\ell) \bar{C}_{TI_j}(\ell) \right] \nonumber\\
 = \:& \sum_j w_{\ell,j} \left[ \bar{C}_{TI_k}(\ell) \bar{C}_{TI_j}(\ell)
 - \bar{C}_{TI_k}(\ell) \bar{C}_{TI_j}(\ell) \right] = 0 \,.
\end{align}
Since the solution that minimizes the difference between $T$ and $\tilde{T}$
also makes $\partial\bar{r}(\ell)/\partial w_{\ell,k}=0$, it also maximizes
the cross-correlation coefficient of the mean power spectrum. However, this
does not guarantee that it maximizes the mean cross-correlation of the power
spectra
\begin{equation}
 \frac{1}{N} \sum_{n=1}^{N} \frac{C^n_{T\tilde{T}}(\ell)}
 {\sqrt{C^n_{TT}(\ell) C^n_{\tilde{T}\tilde{T}}(\ell)}} \,.
\end{equation}

The above calculation assumes that the linear weights are scale dependent,
i.e. there is one set of weights for each $\ell$ bin. One can generalize
the above derivation to the \emph{scale-independent} weights as
\begin{equation}
 \tilde{T} = \sum_i w_i I_i \,.
\end{equation}
In other words, there is only one set of weights for all $\ell$ bins.
In practice we find that the scale-independent weights suffer less
overfitting issue that the scale-dependent weights, hence in this paper
we shall only present the results of the linear combination model
using the scale-independent weights.

\section{Effect of $\lambda$ on the correlation between the predicted
and target maps}
\label{app:lambda}

\begin{figure}[h]
\centering
\includegraphics[width=0.7\textwidth]{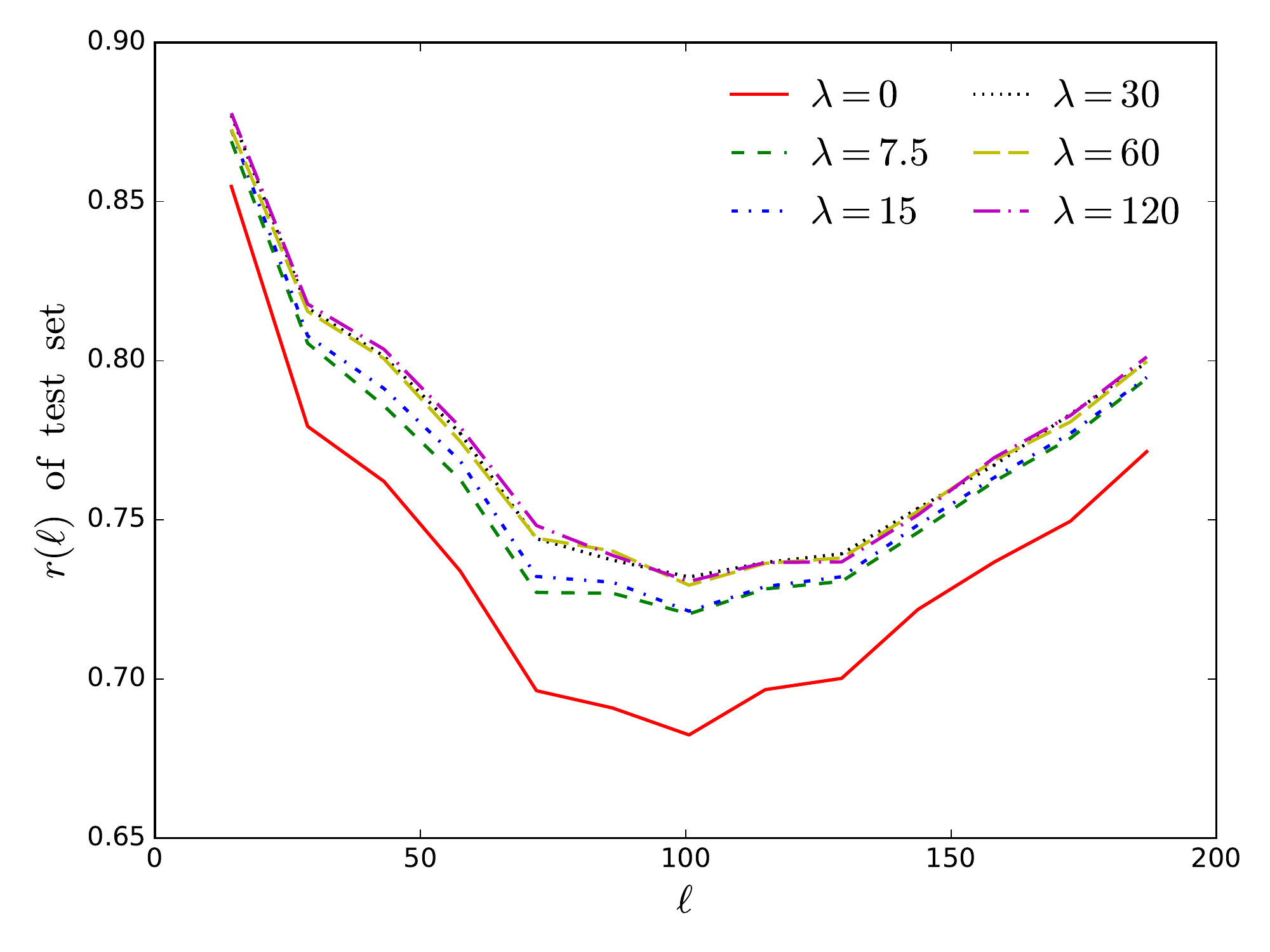}
\caption{Mean cross-correlation coefficient between the target dust maps
and different models as a function of $\ell$ for the test set for various
$\lambda$ values (shown by different colors and line styles).}
\label{fig:rl_lambda}
\end{figure}

The loss function for training the neural network is defined in eq.~\ref{eq:loss}
and $\lambda$ controls the relative contributions between the cross-correlation
and the pixel difference. To study how the value of $\lambda$ affects the results
of the neural network, we train the U-Net model for different $\lambda$ from 0 to
120. Figure~\ref{fig:rl_lambda} presents the mean cross-correlation coefficient
between the target dust maps and different models as a function of $\ell$ for the
test set for various $\lambda$ values (shown by different colors and line styles).
We find that the correlation is lower for smaller $\lambda$. In particular, for
$\lambda=0$ the correlation is significantly lower. This is likely because using
only the pixel difference in the loss function puts too much weight on the small-scale
information and the improvement on the large-scale correlation, which is our main
scientific interest, is limited. As $\lambda\ge30$, there is no significant change
for the correlation on all scales. Thus, in this paper we set $\lambda=30$ and the
two loss terms have similar amplitudes.

\section{Result for east and west galactic hemisphere division}
\label{app:split}
\begin{figure}
\centering
\includegraphics[width=0.495\textwidth]{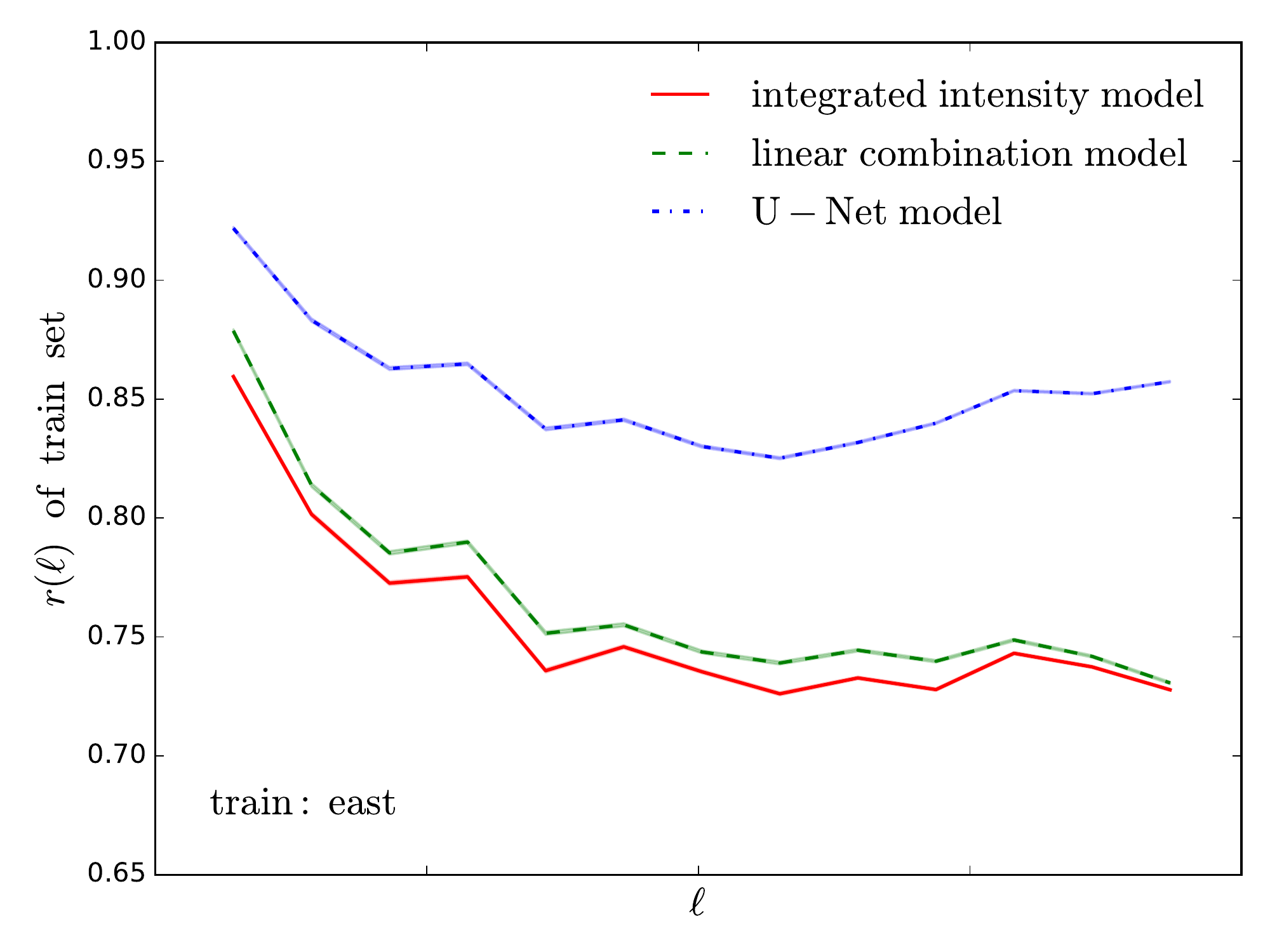}
\includegraphics[width=0.495\textwidth]{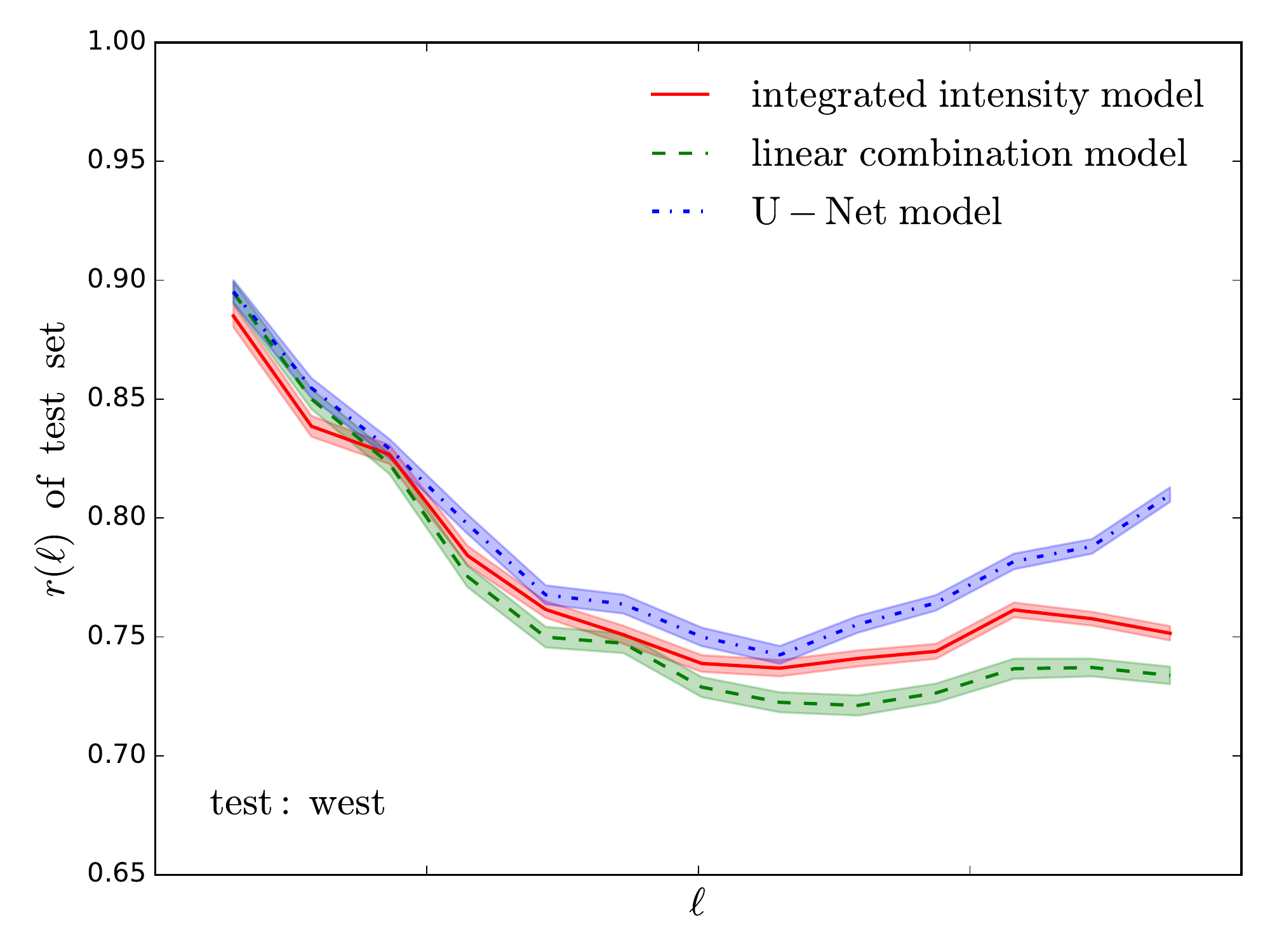}
\caption{Same as figure~\ref{fig:rl}, but the training and test sets are taken
from east and west galactic hemispheres, respectively.}
\label{fig:rl_split2}
\end{figure}

In figure~\ref{fig:rl}, we find a significant difference of the integrated intensity
model performance in training (north) and test (south) sets. To investigate this,
we divide the sky into east and west galactic hemispheres, and the rest of the
data preparation follows the procedures in section~\ref{sec:method}. To ensure no
overlap for data in the two hemispheres, we add a buffer of 36 degrees, so at the
equator the east and west hemisphere has RA ranges of $(198^{\circ}, 342^{\circ})$
and $(18^{\circ}, 162^{\circ})$. We then train the models using the data from the
east hemisphere and access the model performance in the west hemisphere. Figure~\ref{fig:rl_split2}
shows the training and test results in left and right panel, respectively. First,
we note that there is no significant difference of the integrated intensity model
for training and test sets. Thus, the difference we see between north and south
galactic hemispheres is likely a statistical fluctuation. In addition, we find
similar conclusion as in figure~\ref{fig:rl}. Namely, for the test set, the linear
model has a similar performance as the integrated intensity model, whereas the
U-Net model outperforms. However, the improvement is more significant on small
scales. Since a different sky division yields comparable results, we conclude
that the training of the neural network does not suffer strong overfitting.

\bibliographystyle{JHEP}
\bibliography{refs}

\end{document}